\documentclass[]{aa}
\usepackage{graphicx}
\usepackage{color}
\usepackage{txfonts}
\usepackage{amssymb}

\begin{document}


\title{Fast magnetohydrodynamic oscillation of longitudinally inhomogeneous prominence threads: an analogue with quantum harmonic oscillator}

\author{S. N. Lomineishvili\inst{1,4}, T.~V.~Zaqarashvili\inst{1,4}, I.~Zhelyazkov\inst{2}, and A. G. Tevzadze \inst{3}}
 \institute{Space Research Institute, Austrian Academy of Sciences, 8042 Graz, Austria\\
             \email{teimuraz.zaqarashvili@oeaw.ac.at}
                \and
         Faculty of Physics, Sofia University, 5 James Bourchier Blvd., 1164 Sofia, Bulgaria
                \and
      Faculty of Exact and Natural Sciences, Javakhishvili Tbilisi State University,
              1 Chavchavadze Avenue, 0128 Tbilisi, Georgia
              \and
         Abastumani Astrophysical Observatory at Ilia State University, 3/5 Cholokashvili Avenue, 0162 Tbilisi, Georgia\\}

\date{Received / Accepted }

\abstract
{Previous works indicate that the frequency ratio of second and first harmonics of kink oscillations has tendency towards 3 in the case of prominence threads. This is not a straightforward result, therefore it requires adequate explanation.}
{We aim to study the magnetohydrodynamic oscillations of longitudinally inhomogeneous
prominence threads and to shed light on the problem of frequency ratio.}
{Classical Sturm--Liouville problem is used for the threads with longitudinally inhomogeneous plasma density. We show that the spatial variation of total pressure perturbations along the thread is governed by the stationary Schr\"{o}dinger equation, where the longitudinal inhomogeneity of plasma density stands for the potential energy. The Schr\"{o}dinger equation appears as the equation of quantum harmonic oscillator for a parabolic profile of plasma density. Consequently, the equation has bounded solutions in terms of Hermite polynomials. }
{Boundary conditions at the thread surface lead to transcendental dispersion equation with Bessel functions. Thin flux tube approximation of the dispersion equation shows that the
frequency of kink waves is proportional to the expression $\alpha(2n+1)$, where $\alpha$ is the density inhomogeneity parameter and $n$ is the longitudinal mode number. Consequently,  the ratio of the frequencies of second and first harmonics tends to $3$ in prominence threads.  Numerical solution of the dispersion equation shows that the ratio only slightly decreases for thicker tubes in the case of smaller longitudinal inhomogeneity of external density, therefore the thin flux tube limit is a good approximation for prominence oscillations. However, stronger longitudinal inhomogeneity of external density may lead to the significant shift of frequency ratio for wider tubes and therefore the thin tube approximation may fail.}
{The tendency of frequency ratio of second and first harmonics towards 3 in prominence threads is explained by the analogy of the oscillations with quantum harmonic oscillator, where the density inhomogeneity of the threads plays a role of potential energy.}


\titlerunning{MHD oscillations of prominence threads}

\authorrunning{Lomineishvili et al.}

\maketitle


\section{Introduction}
     \label{S-Introduction}

Filaments or prominences are cold clouds of dense plasma imbedded in the tenuous hot solar corona, probably supported
by the prominence magnetic field against the gravity. Observations show that prominences are made up of many tube-like fine structures very likely smaller than the currently achievable resolution ($\approx\!\!\!150$~km). The structures are called fibrils or threads and they have a length of 10$^3$--10$^4$~km, which is much shorter than the longitudinal extent of prominence itself (Lin et al. \cite{Lin2005}, Labrosse et al. \cite{Labrosse2010}). Therefore, it is likely that the threads are concentrations of cold plasma at the middle of much longer magnetic tube, while the remaining part of the tube is filled with hot coronal plasma.

Small-amplitude magnetohydrodynamic (MHD) waves and oscillations are frequently observed in solar
prominence threads (Oliver et al. \cite{Oliver2002}, Lin et al. \cite {Lin2005}, Lin et al. \cite {Lin2007}, Lin et al. \cite{Lin2009}, Mackay et al. \cite {Mackay2010}, Arregui et al. \cite {Arregui2012}). MHD oscillations in
prominences are well studied by both, slab and tube approximations
(Roberts \cite{Roberts1991}, Joarder and Roberts \cite{Joarder1992a,Joarder1992b}, Oliver et al. \cite{Oliver1993}, Joarder et al. \cite{Joarder1997}, D\'iaz et al. \cite{Diaz2002,Diaz2005}, Dymova and Ruderman \cite{Dymova2005}, Terradas et al. \cite{Terradas2008}, Oliver \cite{Oliver2009}, Arregui et al. \cite{Arregui2011,Arregui2012}).
It turns out that the longitudinal density stratification  significantly influences the oscillation spectra of simple homogeneous tubes (D\'iaz et al. \cite{Diaz2004,Diaz2010}, Andries et al. \cite{andries1,Andries2009}, Dymova and Ruderman \cite{Dymova2005}, McEwan et al. \cite{McEwan2006}, Zaqarashvili et al. \cite{Zaqarashvili20072,Zaqarashvili2013}).

However, the oscillations in coronal loops and prominence threads are very different. The difference is caused by the density profile: coronal loops have denser plasma at footpoints probably due to the stratification, while prominence threads (fibrils) are denser at the midpoint of magnetic tube (or slab). The density contrast between  the tube center and footpoints is by an order larger in prominences than in coronal loops. The different physical parameters lead to different oscillation spectra in these two structures. For example, the ratio
between the periods of first and second harmonics in homogeneous coronal loops is near 2 (Edwin and Roberts \cite{Edwin1983}), while the stratification may lead to the significant shift of the ratio (Andries et al. \cite{andries2}, McEwan et al. \cite{McEwan2006}). There are several observational examples of simultaneous existence of first and second harmonics of MHD oscillations in coronal loops, which indeed show the deviation of the ratio from 2 (Van Doorsselaere et al. \cite{van1, Van2}, Verwichte et al. \cite{Verwichte}, De Moortel and Brady \cite{demoo}, Srivastava et al. \cite{Srivastava, Srivastava2013}, Inglis and Nakariakov \cite{Inglis}).

On the other hand, the period ratio of first and second harmonics is not 2 in the prominence case. Early calculations showed that the asymptotic frequency of internal kink mode in the prominence case is proportional to $2n+1$, where $n$ is the mode number (Joarder and Roberts \cite{Joarder1992b}). Then, the period ratio of first and second harmonics is 3, in contrast with coronal loops. The similar ratio can be seen on the plots of other papers concerning the prominence oscillations (Dymova and Ruderman \cite{Dymova2005}, D\'iaz et al. \cite{Diaz2010}). However, the authors did not clearly note this fact and consequently did not explain why the period ratio has tendency towards 3 in prominences. On the other hand, the frequency dependence of $2n+1$ found by Joarder and Roberts (\cite{Joarder1992b}) suggests that the quantum mechanical analogy may stand behind the result. All these authors used piecewise profiles to study the prominence oscillations. Consequently, the concentration of density in the middle of tube can be considered as a potential energy, which may lead to the problem of quantum harmonic oscillator.

In this paper, we reconsider the oscillation of prominence threads using parabolic density profile in the longitudinal direction. Solution of Sturm-Liouville problem  allows us to obtain the equation of a quantum harmonic oscillator. The solution of the equation leads to the frequency dependence of $2n+1$, therefore, we may conclude that the enhancement of plasma density (both, piecewise and parabolic profiles) at the middle of flux tube allows to quantify waves analogously with quantum mechanics.

The paper is organized as follows: In Sect. 2 we describe the physical
model of the considered problem. In Sect. 3 we give the Sturm-Liouville
solution that leads to the dispersion equation described in Sect. 4.
Analytical and numerical solutions of the dispersion
equation are given in Sects. 4.1 and 4.2. We shortly summarize the
paper in Sect. 5.

\section{Main equations} 
      \label{S-general}

We consider a prominence thread as a magnetic tube of length $2L$ and radius $a$ in the
cylindrical coordinate system $(r,\varphi,z)$. We assume that the unperturbed density inside the tube is homogenous along radial direction and
has longitudinal parabolic inhomogeneity along the z-axis in the form of
\begin{equation}\label{rhoi}
{\rho_0}(r<a)={\rho_{\rm i}}={\rho_{\rm i0}\left(1-\alpha^2{z^2\over
L^2}\right )},
\end{equation}
where $\alpha$ is a constant and $\rho_{\rm i0}$ is the unperturbed density at the tube midpoint (i.e., at $z=0$) inside the thread, which has a typical value of prominences. The thread is imbedded in a coronal plasma with much smaller density. The density is higher at the tube midpoint and decreases towards the
ends approaching to its coronal value (Fig.~1). The thread itself probably is a part of much longer magnetic tube, which has a coronal density outside the thread.
Magnetic field is uniform and longitudinal, i.e., the tube is non-twisted.


We consider cold plasma approximation without gravity, then the single-fluid linear ideal MHD equations can be written in cylindrical coordinates as
\begin{equation}\label{linear-r}
{\partial u_r\over\partial t}={B_0\over 4\pi\rho_0}{\partial b_r\over \partial
z}-{B_0\over 4\pi\rho_0}{\partial
b_z\over\partial r},
\end{equation}
\begin{equation}\label{linear-phi}
{\partial u_\varphi\over\partial t}={B_0\over 4\pi\rho_0}{\partial
b_\varphi\over\partial z}-{B_0\over 4\pi\rho_0 r}{\partial
b_z\over\partial\varphi},
\end{equation}
\begin{equation}\label{linear-br}
{\partial b_r\over\partial t}=B_0{\partial u_r\over\partial z},
\end{equation}
\begin{equation}\label{linear-bphi}
{\partial b_\varphi\over\partial t}=B_0{\partial
u_\varphi\over\partial z},
\end{equation}
\begin{equation}\label{linear-bz}
{\partial b_z\over\partial t}=-B_0{u_r\over r}-B_0{\partial
u_r\over\partial r}-{B_0\over r}{\partial
u_\varphi\over\partial\varphi},
\end{equation}
where $\rho_0$ is the unperturbed density, $B_0$ is the unperturbed
magnetic field and $b_r$, $b_\varphi$, $b_z$, $u_r$, $u_\varphi$ are
magnetic field and velocity perturbations, respectively. Note that
the longitudinal velocity component is zero, i.e., $u_z=0$, in
$\beta \; (=\!\!4\pi p/B^2) = 0$ approximation.
Equations~(\ref{linear-r})--(\ref{linear-bz}) represent a closed set
of equations, therefore the continuity equation is not necessary
here. Density perturbations can be expressed by magnetic field
perturbations later, if necessary.

Equations~(\ref{linear-r})--(\ref{linear-bz}) lead, after some
algebra, to the following equation
\begin{equation}\label{general}
{1 \over r}{\partial\over\partial r} \left(r{\partial p_{\rm m}\over
\partial r}\right)+{1 \over r^2}{\partial^2 p_{\rm m}\over
\partial\varphi^2}+{\partial^2 p_{\rm m}\over\partial z^2}- {1\over
v_{\rm A}^2}{\partial^2 p_{\rm m}\over\partial t^2}=0,
\end{equation}
where
\begin{equation}\label{magnetic-pressure}
p_{\rm m}={B_0 b_z \over 4\pi}
\end{equation}
is the perturbed magnetic pressure and
\begin{equation}\label{alfven speed}
v_{\rm A}={B_0\over\sqrt{4\pi\rho_0}}
\end{equation}
is the Alfv\'en speed.

In order to find the spectrum of MHD oscillations in a magnetic
tube, one should solve Eq.~(\ref{general}) inside and outside the
tube and then merge the solutions at the tube surface through
continuity of the total pressure and radial velocity.


\section{Sturm-Liouville problem and Schr\"{o}dinger
equation}

Fourier analysis of Eq.~(\ref{general}) with $\exp(-\mathrm{i}\omega
t+\mathrm{i}m \varphi)$ using Eq.~(\ref{rhoi}) leads to the following equation inside the
tube
\begin{equation}\label{eq-inside}
{1\over r}{\partial\over\partial r}\left(r{\partial p_{\rm m}\over
\partial r}\right)+{\partial^2 p_{\rm m}\over\partial z^2} +
\left [{\omega^2\over v_{\rm Ai0}^2 }\left (1- {{\alpha^2 z^2}\over
L^2}\right ) - {m^2\over  r^2}\right ]p_{\rm m}=0,
\end{equation}
where
\begin{equation}\label{alfvenspeed-inside}
v_{\rm Ai0}={B_0\over\sqrt{4\pi\rho_{\rm i0}}}
\end{equation}
is the Alfv\'en speed at $z=0$.

\begin{figure}
\vspace*{1mm}
\begin{center}
\includegraphics[width=9.5cm]{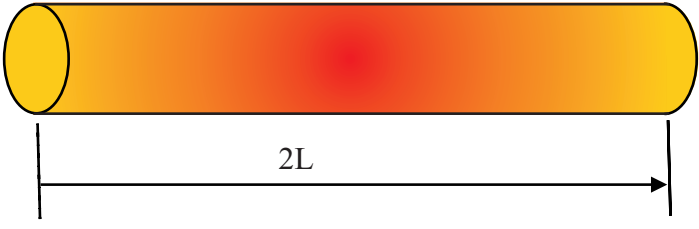}
\end{center}
\caption{Prominence thread with a longitudinal inhomogeneity of plasma density.}
\end{figure}

We assume that the density outside the tube has the parabolic profile along  $z$-axis as inside the tube, namely
\begin{equation}\label{rhoe}
{\rho_0}(r>a)={\rho_{\rm e}}={\rho_{\rm e0}\left(1-\alpha_{\rm e}^2{z^2\over L^2}\right )},
\end{equation}
where $\alpha_{\rm e}$ is a constant and $\rho_{\rm e0}$ is unperturbed density at $z=0$ outside the tube. Then the equation similar to Equation~(\ref{eq-inside}) is obtained outside the tube
\begin{equation}\label{eq-outside}
{1\over r}{\partial\over\partial r}\left(r{\partial p_{\rm m}\over
\partial r}\right)+{\partial^2 p_{\rm m}\over\partial z^2} +
\left [{\omega^2\over v_{\rm Ae0}^2 }\left (1- {{\alpha_{\rm e}^2 z^2}\over
L^2}\right ) - {m^2\over  r^2}\right ]p_{\rm
m}=0,
\end{equation}
where
\begin{equation}\label{alfvenspeed-outside}
v_{\rm Ae0}={B_0 \over\sqrt{4\pi \rho_{\rm e0}}}.
\end{equation}

Equations~(\ref{eq-inside}) and  (\ref{eq-outside}) with
corresponding boundary conditions represent a classical
Sturm-Liouville problem and they can be solved by the method of
separation of variables.

We assume
\begin{equation}\label{separation}
p_{\rm m}(r,z)=p_{\mathrm{m}r}(r)p_{\mathrm{m}z}(z).
\end{equation}

Then Eq.~(\ref{eq-inside}) leads to the following equations
\begin{equation}\label{eq-inside-r}
{1\over r}{\partial\over\partial r}\left(r{\partial p_{{\rm m}r}\over
\partial r}\right)+\left({\omega^2\over v_{\rm Ai0}^2} -
{m^2 \over r^2}\right)p_{{\rm m}r}=k_n^2 p_{{\rm m}r},
\end{equation}
\begin{equation}\label{eq-inside-z}
{\partial^2 p_{{\rm m}z}\over\partial z^2 }- {{\omega^2\over v_{\rm
Ai0}^2}\alpha^2{z^2\over L^2}p_{{\rm m}z}}=-k_n^2 p_{{\rm m}z},
\end{equation}
where $k_n^2$ is the separation constant.

Equation~(\ref{eq-inside-z}) is analogous to the stationary Schr\"{o}dinger
equation, where the first and second terms in the left hand side
correspond to the kinetic and potential energies (after
corresponding normalization), while the term in the right hand side
corresponds to the total energy. More specifically, with substitution of
\begin{equation}\label{eq-os}
\alpha/(v_{\rm Ai0}L)=m/\hbar,\,k_n=\sqrt{2mE}/\hbar,\,p_{{\rm m}z}=\Psi(z),
\end{equation}
Eq.~(\ref{eq-inside-z}) is rewritten as
\begin{equation}\label{eq-osc}
{{-{\hbar}^2}\over {2m}}{\partial^2 \Psi(z)\over\partial z^2 }+{1\over 2}m\omega^2z^2\Psi(z) =E\Psi(z),
\end{equation}
which is the equation of a quantum harmonic oscillator (Griffiths \cite{Griffiths2004}). It must be noted that the considered problem is entirely classical: the Planck's constant is introduced here just to demonstrate the mathematical similarity of the considered problem with the behaviour of a quantum particle in a potential well. This equation has exact solutions and governs many dynamical systems in different branches of science.

Assuming $\alpha=0$ in Eq.~(\ref{eq-inside-z}) leads to the solution of homogeneous tube and after corresponding calculation one may recover well known dispersion relations obtained by Edwin and Roberts (\cite{Edwin1983}).

Now we introduce a new variable
\begin{equation}\label{variable}
x \equiv z\sqrt{2\alpha\omega\over{v_{\rm Ai0}L}},
\end{equation}
then, from Eq.~(\ref{eq-inside-z}) we obtain the equation of
parabolic cylinder
\begin{equation}\label{weber}
{\partial^2 p_{{\rm m}z}\over\partial x^2}- \left({x^2\over 4}+d\right
)p_{{\rm m}z}=0,
\end{equation}
where
\begin{equation}\label{d}
d \equiv-{v_{\rm Ai0}L\over{2\alpha\omega}}k_n^2<0,
\end{equation}
for $\omega > 0$. For $\omega < 0$, we have $d>0$.

After the substitution of Eq.~(\ref{variable}), it is no longer possible to recover the dispersion relation of homogeneous tube. The approximation of homogeneous tube means $\alpha=0$, which implies $x=0$ everywhere and consequently  Eq.~(\ref{weber}) is not determined.

The solutions of Eq.~(\ref{weber}) are parabolic cylinder or Weber
functions $U(d,x)$ and $V(d,x)$ (Abramowitz and Stegun \cite{abramowitz}).


When
\begin{equation}\label{dn}
d=-n-{1\over 2},
\end{equation}
where $n=0,1,2,\ldots$, the bounded solutions can be expressed in terms of
Hermite polynomials (Abramowitz and Stegun \cite{abramowitz}):
\begin{equation}\label{U}
U\left(-n-{1\over 2},x\right )=2^{-{n\over 2}}\mathrm{e}^{-{{x^2}\over
4}}H_n\left ({x\over{\sqrt{2}}}\right ),
\end{equation}
for $\omega>0$ and
\begin{equation}\label{V}
V\left(n+{1\over 2},x\right )=2^{-{n\over 2}}\mathrm{e}^{{x^2}\over
4}(-\mathrm{i})^nH_n\left({\mathrm{i}x\over{\sqrt{2}}}\right ),
\end{equation}
for $\omega<0$. The Hermite polynomials are set of orthogonal
polynomials over the domain $(-\infty, \infty)$ with weighting
function $\exp(x^2/2)$, therefore the functions $U\left(-n-{1\over
2},x\right )$ and $V\left(n+{1\over 2},x\right )$ form an orthogonal
basis of the Hilbert space.

Equations~(\ref{d}) and (\ref{dn}) imply that
\begin{equation}\label{k_n}
k^2_n={{\alpha\omega}\over{v_{\rm Ai0}L}}(2n+1).
\end{equation}

Equations ~(\ref{eq-os}) and (\ref{k_n}) lead to the equation
\begin{equation}\label{fund}
E_n = \hbar \omega_n \left (n+{1\over 2}\right ),
\end{equation}
which is well known formula of energy levels and thus the oscillation of prominence threads is analogous with quantum harmonic oscillator. Nakariakov and Oraevsky (\cite{Nakariakov}) also studied a plasma non-uniformity with a parabolic
(transverse) profile in the context of MHD oscillations of coronal plasma structures and the solutions of a boundary problem in terms of the Hermite polynomials with the use of
the quantum mechanical analogy were found.

From Eq.~(\ref{k_n}) it is seen that $\omega<0$ leads to
imaginary $k_n$, therefore the solution can be neglected due to
physical reasons. Therefore, in the future we only consider the
solution with $\omega>0$, i.e., Eq.~(\ref{U}). This solution rapidly decreases along $x$ due to the factor of $\mathrm{e}^{-{{x^2}/
4}}$, therefore it is bounded in the longitudinal direction. Hence, here is no need to fix close boundary conditions at the end of much longer tube as it is usually done for piecewise profiles (Dymova and Ruderman \cite{Dymova2005}, D\'iaz et al. \cite{Diaz2010}). Density inhomogeneity of prominence thread traps bounded oscillations analogously with quantum harmonic oscillator.

The general solution of Eq.~(\ref{weber}) is
\begin{equation}\label{general solution}
p_{{\rm m}z}=p_{{\rm m}z}(0) 2^{-{n\over 2}}\mathrm{e}^{-{{x^2}\over
4}}H_n\left({x\over {\sqrt{2}}}\right),
\end{equation}
where $p_{{\rm m}z}(0)$ is a constant.

Now, Eq.~(\ref{eq-inside-r}) can be
rewritten as

\begin{equation}\label{Bessel-equation}
{1\over r}{\partial\over\partial r}\left(r{\partial p_{{\rm m}r{\rm
i}}\over\partial r}\right) + \left[m^2_{\rm i}-{m^2\over
r^2}\right]p_{{\rm m}r{\rm i}}=0,
\end{equation}
where $p_{{\rm m}r{\rm i}}$ is the
radially-dependent part of total pressure perturbations inside the tube and
\begin{equation}\label{m_i}
m^2_{\rm i} ={{\omega}^2\over{v^2_{\rm Ai0}}}-{{\alpha \omega}\over {v_{\rm Ai0}L}}(2n+1).
\end{equation}

Repeating the same calculation outside the tube, we obtain the equation
\begin{equation}\label{Mod-Bessel-equation}
{1\over r}{\partial\over\partial r}\left(r{\partial p_{{\rm m}r{\rm
e}}\over\partial r}\right) - \left[m^2_{\rm e} +{m^2\over
r^2}\right]p_{{\rm m}r{\rm e}}=0,
\end{equation}
where $p_{{\rm m}r{\rm e}}$ is the
radially-dependent part of total pressure perturbations outside the tube, and
\begin{equation}\label{m_e}
m^2_{\rm e} ={{\alpha_{\rm e} \omega}\over {v_{\rm Ae0}L}}(2n+1)-{{\omega}^2\over{v^2_{\rm Ae0}}}.
\end{equation}
Equations~(\ref{Bessel-equation}) and (\ref{Mod-Bessel-equation})
are Bessel and modified Bessel equations, respectively. Therefore, the corresponding solutions are:
\begin{equation}\label{sollution-generali}
p_{{\rm m}r{\rm i}}=a_1J_m(m_{\rm i} r)+a_2Y_m(m_{\rm i} r),
\end{equation}
\begin{equation}\label{sollution-generale}
p_{{\rm m}r{\rm e}}=b_1I_m(m_{\rm e} r)+b_2K_m(m_{\rm e} r),
\end{equation}
where $a_1, a_2, b_1, b_2$ are constants.

Total pressure perturbation inside the tube, $p_{{\rm m}r{\rm i}}$, must be
finite at $r=0$, which yields $a_2=0$. Analogously, total pressure
perturbation outside the tube $p_{{\rm m}r{\rm e}}$ must be finite at $r \to
\infty$, which yields  $b_1=0$. These two conditions lead to the
following expressions of total pressure perturbations inside and
outside the tube:
\begin{equation}\label{sollution-i}
p_{{\rm m}r{\rm i}}=a_1J_m(m_{\rm i} r),
\end{equation}
\begin{equation}\label{sollution-e}
p_{{\rm m}r{\rm e}}=b_2K_m(m_{\rm e} r).
\end{equation}

\section{Dispersion equation}

Dispersion equation for MHD oscillations in the magnetic tube can be
obtained through the boundary conditions at the tube surface. The
boundary conditions require the continuity of total pressure and
radial velocity at $r=a$. Total pressure balance condition at the
tube surface yields:
\begin{equation}\label{pressure balance}
[p_{{\rm m}r{\rm i}}]_a = [p_{{\rm m}r{\rm e}}]_a\Rightarrow
a_1J_m(m_{\rm i} a)=b_2K_m(m_{\rm e} a).
\end{equation}

\begin{figure}
\vspace*{1mm}
\begin{center}
\includegraphics[width=9.5cm]{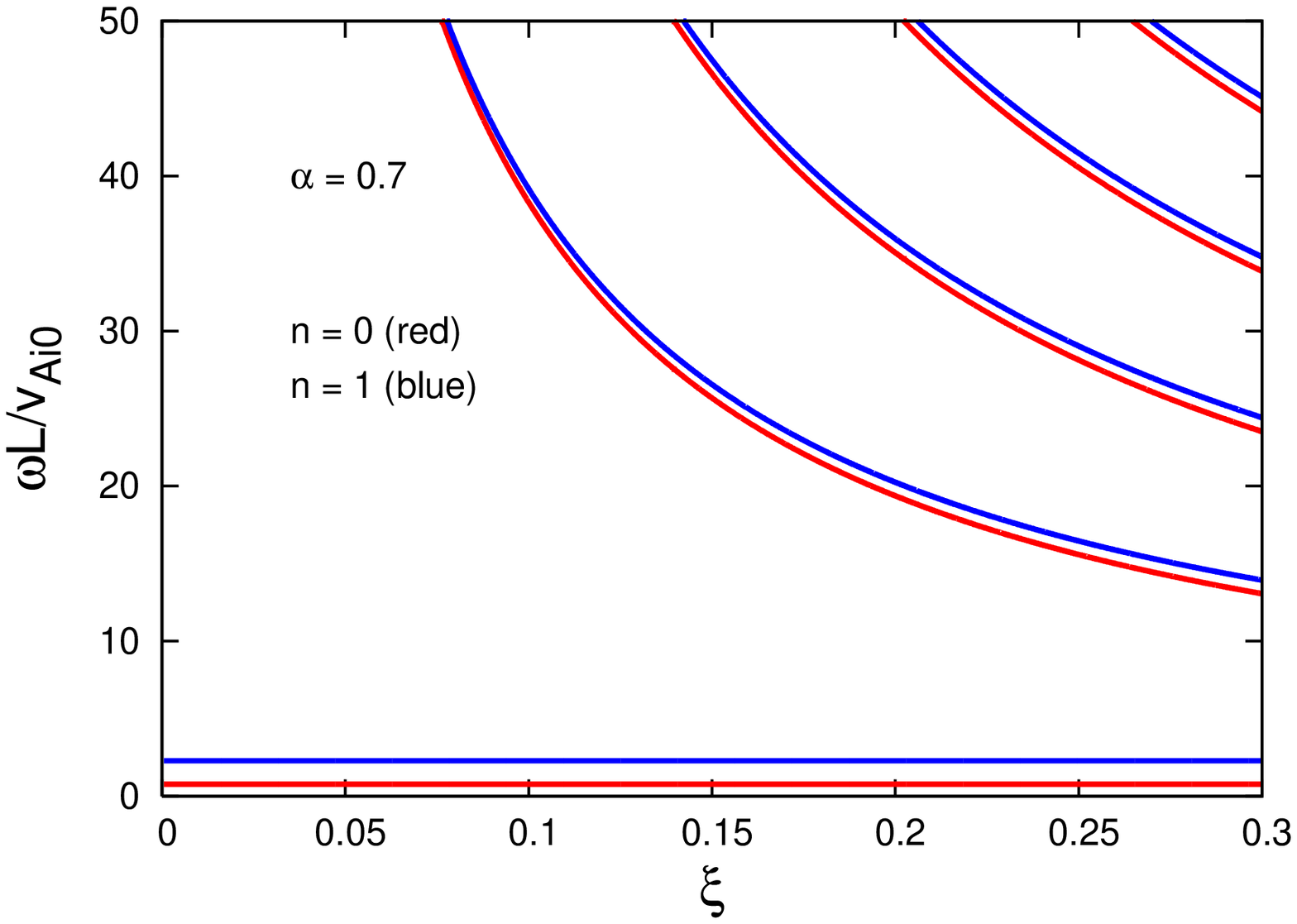}
\includegraphics[width=9.5cm]{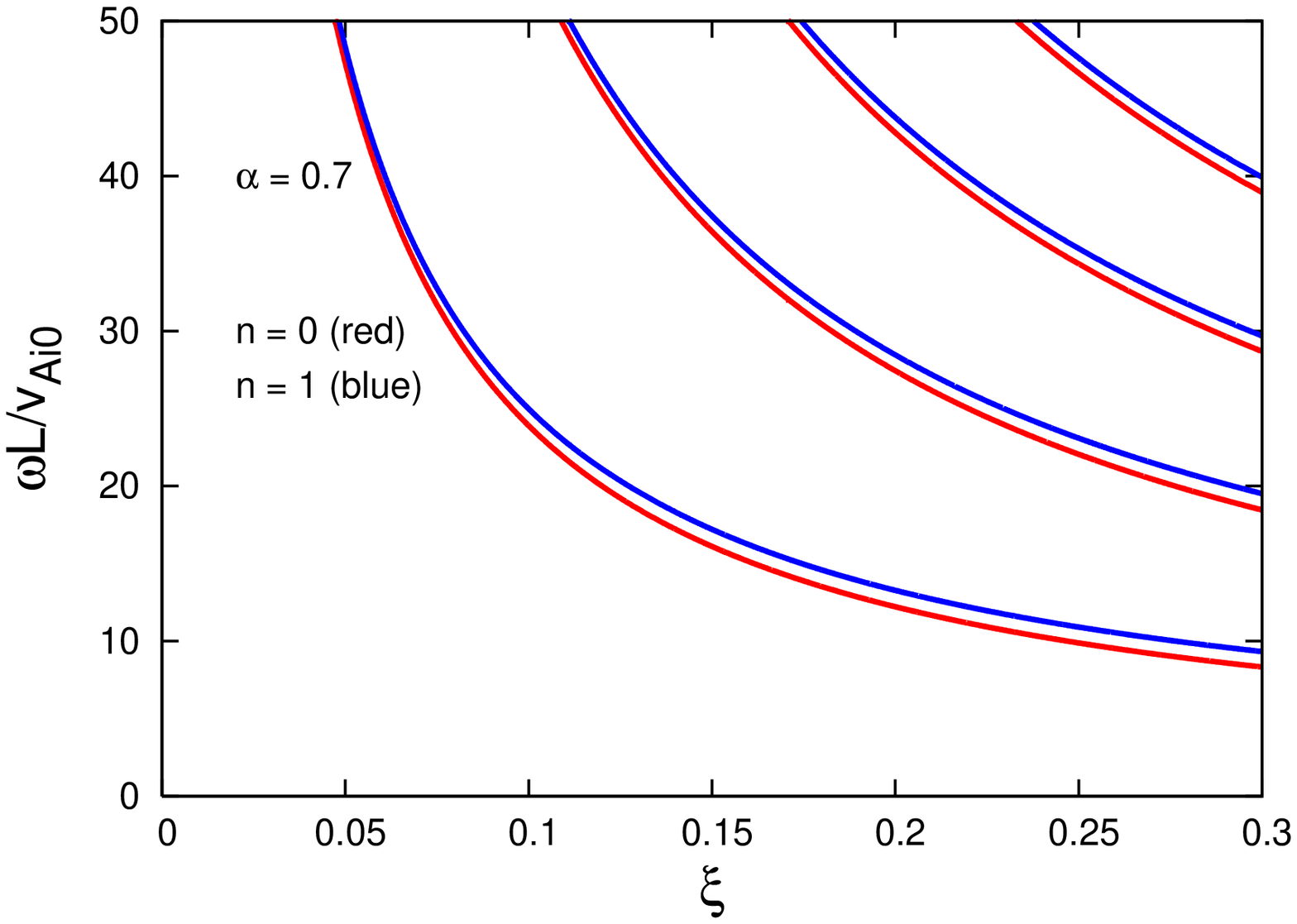}
\end{center}
\caption{Normalized frequency, ${\tilde \omega}={{\omega L}/{v_{\rm
Ai0}}}$, vs the ratio of the tube width to length, $\xi=a/L$, for kink
(upper panel) and sausage (lower panel) waves. $n=0$ (red lines) and
$n=1$ (blue lines) correspond to the first and second harmonics.
Here $\eta=0.01$ and $\alpha=\alpha_{\rm e}=0.7$ are used.}
\end{figure}

The equations for radial velocity perturbations inside the tube can be obtained from Eq.~(\ref{linear-r}) as
\begin{equation}\label{eq-uri}
{\partial^2 u_{r{\rm i}}\over\partial z^2} + {{\omega^2 }\over
{v^2_{\rm Ai}}}u_{r{\rm i}} =-{{\mathrm{i}\omega}\over{\rho_{\rm
i}v^2_{\rm Ai}}}{\partial p_{\rm mi}\over\partial r},
\end{equation}

Next we assume
\begin{equation}\label{uri}
u_{r{\rm i}}(r,z)=u_{r{\rm i}}(r)p_{{\rm m}z}(z),
\end{equation}
then, Eq.~(\ref{eq-uri}) leads to the following
equation:
\begin{equation}\label{eq1-uri}
u_{r{\rm i}}(r)\left[{\partial^2 p_{{\rm m}z}(z)\over \partial
z^2}+{{\omega^2 }\over {v^2_{\rm Ai}}}p_{{\rm m}z}(z)\right]=
-{{\mathrm{i}\omega p_{{\rm m}z}(z)}\over{\rho_{\rm i}v^2_{\rm Ai}}}
{\partial p_{{\rm m}r{\rm i}}(r)\over\partial r}.
\end{equation}

\begin{figure}
\vspace*{1mm}
\begin{center}
\includegraphics[width=9.5cm]{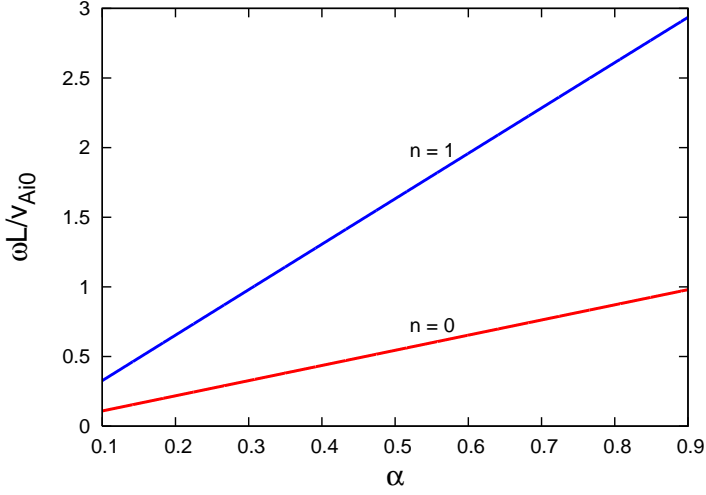}
\end{center}
\caption{Normalized frequency of first (red curve) and second
(blue curve) harmonics of fundamental kink waves vs the
inhomogeneity parameter, $\alpha$, for $\eta=0.01$\textbf{,} in the case of $\alpha_{\rm e}=\alpha$.}
\end{figure}

Using Eq.~(\ref{eq-inside-z}) this equation can be rewritten after
some algebra as
\begin{equation}\label{uri1}
u_{r{\rm i}}(r)= -{{\mathrm{i}\omega\over{\rho_{\rm
i0}(\omega^2-{v^2_{\rm Ai0}}k^2_n)}}{\partial p_{{\rm m}r{\rm
i}}(r)\over
\partial r}}.
\end{equation}

Similar calculations outside the tube leads to the equation
\begin{equation}\label{ure1}
u_{r{\rm e}}(r)= -{{\mathrm{i}\omega\over{\rho_{\rm
e0}(\omega^2-{v^2_{\rm Ae0}}k^2_{n{\rm e}})}}{\partial p_{{\rm m}r{\rm
e}}(r)\over
\partial r}},
\end{equation}

where
\begin{equation}\label{k_ne}
k^2_{n{\rm e}}={{\alpha_{\rm e}\omega}\over{v_{\rm Ae0}L}}(2n+1).
\end{equation}

Continuity of transverse velocity at the tube surface yields
\begin{equation}\label{cont-velocity}
\rho_{\rm e0}(\omega^2-{v^2_{\rm Ae0}}k^2_{n{\rm e}})\left [{{\partial
p_{{\rm m}r{\rm i}}\over\partial r}}\right ]_{a}= \rho_{\rm
i0}(\omega^2-{v^2_{\rm Ai0}}k^2_n)\left [{{\partial p_{{\rm m}r{\rm
e}}\over\partial r}}\right ]_{a}.
\end{equation}
With the help of Eqs.~(\ref{sollution-i})--(\ref{pressure balance})
we obtain the final transcendental dispersion equation
\begin{equation}\label{trans}
R_{\rm e0}{J^{\prime}_m(R_{\rm i0})\over J_m(R_{\rm i0})} = -R_{\rm
i0}{K^{\prime}_m(R_{\rm e0})\over K_m(R_{\rm e0})},
\end{equation}
where the prime sign, $^{\prime}$, indicates a differentiation by
Bessel function argument and
\begin{equation}\label{Rio}
R_{\rm i0}=\xi \sqrt{{\tilde \omega}^2-{\alpha}{\tilde \omega}(2n+1)},
\end{equation}
\begin{equation}\label{Reo}
R_{\rm e0}=\xi \sqrt{{\alpha_{\rm e}}{\tilde \omega}\sqrt{\eta}(2n+1)-\eta {\tilde
\omega}^2},
\end{equation}
where
\begin{equation}
\xi={a\over L},\,\,\,{\tilde \omega}={{\omega L}\over{v_{\rm
Ai0}}},\,\,\, \eta={\rho_{\rm e0}\over \rho_{\rm i0}}.
\end{equation}
Here $m=0$ and $m=1$ correspond to the sausage and kink waves
respectively, while $n=0$ ($n=1$) is the first (second) harmonic.

\subsection{Analytical solution}

The thin flux tube approximation, i.e., $\xi={a/L}\ll 1$ yields
$R_{\rm i0}\ll 1$ and $R_{\rm e0} \ll 1$. Then, using the recurrence
relations of the Bessel functions,
$J^{\prime}_m(z)=-J_{m+1}(z)+(m/z)J_m(z)$ and
$K^{\prime}_m(z)=-K_{m+1}(z)+(m/z)K_m(z)$, and the expansion of
Bessel functions for small arguments (Abramowitz and Stegun \cite{abramowitz})
\begin{equation}\label{J}
J_m(z)\sim \left ({z\over 2} \right )^m{1\over \Gamma(m+1)},\,\,m
\neq -1,-2,-3, \ldots,
\end{equation}
\begin{equation}\label{K}
K_m(z)\sim {1\over 2} \Gamma(m)\left ({z\over 2} \right )^{-m},\,\,
\mathrm{Re}(m) > 0,
\end{equation}
Equation ~(\ref{trans}) can be approximated for kink waves ($m=1$) as
$$
\eta \xi^2 {\tilde \omega_n}^3-\xi^2(2n+1)(\alpha_{\rm e}\sqrt{\eta}+\alpha\eta){\tilde\omega_n}^2+[\xi^2\alpha\alpha_{\rm e}\sqrt{\eta}(2n+1)^2
\nonumber
$$
\begin{equation}
-4\eta-4]{\tilde\omega_n}+4(2n+1)(\alpha+\alpha_{\rm e}\sqrt{\eta})=0.
\end{equation}

\begin{figure}
\vspace*{1mm}
\begin{center}
\includegraphics[width=9.5cm]{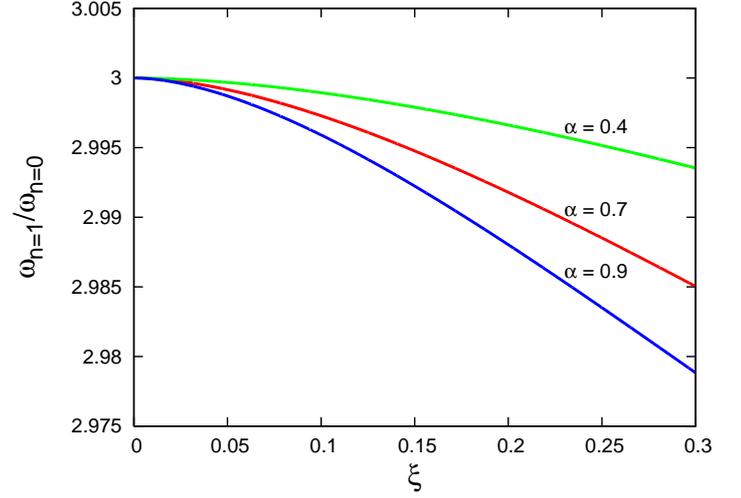}
\end{center}
\caption{Frequency ratio, $\omega_{n=1}/\omega_{n=0}$, of the kink
mode vs $\xi$ for $\eta=0.01$ and three different values of $\alpha$
equal to 0.4 (green line), 0.7 (red line) and 0.9 (blue line)\textbf{,} in the case of $\alpha_{\rm e}=\alpha$.}
\end{figure}

From this equation we obtain the following expression for the
fundamental kink wave (using $\xi \ll 1$)
\begin{equation}\label{fund}
{\omega_n}\approx {{v_{\rm
Ai0}}\over {L}}(2n+1){{\alpha+\alpha_{\rm e}\sqrt{\eta}}\over {1+\eta}}.
\end{equation}

Equation (\ref{fund}) shows that the factor $2n+1$ appears in the expression of kink wave frequency. As we expected this is due to the fact that the density enhancement at the tube midpoint stands for the potential energy, therefore the oscillations are similar to the case of quantum harmonic oscillator. It is also seen that the frequency of kink waves is proportional to the inhomogeneity parameters $\alpha$ and $\alpha_{\rm e}$, which stand for the depth of the potential well. The first harmonic or basis energy level of the oscillations is then
\begin{equation}\label{first}
{\omega_0}\approx {{v_{\rm
Ai0}}\over {L}}{{\alpha+\alpha_{\rm e}\sqrt{\eta}}\over {1+\eta}},
\end{equation}
while the second harmonic or second energy level is
\begin{equation}\label{second}
{\omega_1}\approx {{3v_{\rm
Ai0}}\over {L}}{{\alpha+\alpha_{\rm e}\sqrt{\eta}}\over {1+\eta}}.
\end{equation}

Then the ratio of the frequencies of second and first harmonics is
\begin{equation}\label{ratio}
{{\omega_1}\over {\omega_0}}\approx 3.
\end{equation}
This is the ratio, which has been obtained from time to time in the case of prominence oscillations without physical explanation (Joarder and Roberts \cite{Joarder1992b}, Dymova and Ruderman \cite{Dymova2005}, D\'iaz et al. \cite{Diaz2010}).


To check the analytical solution of the dispersion relation, we use numerical solutions of Eq.~(\ref{trans}), which are presented in the next subsection.

\begin{figure}
\vspace*{1mm}
\begin{center}
\includegraphics[width=9.5cm]{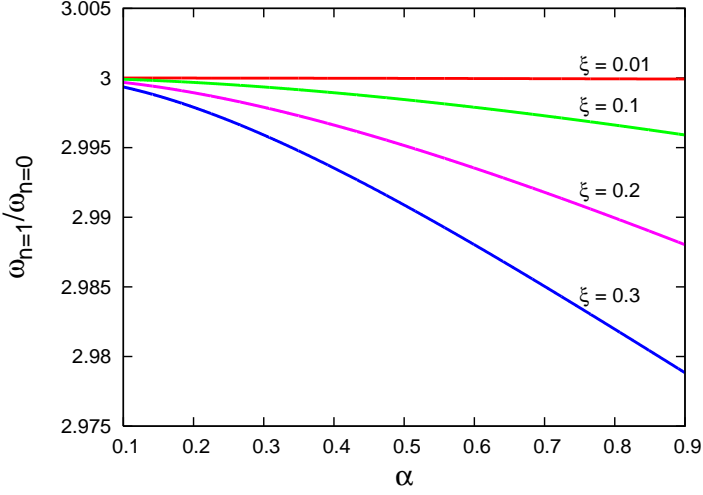}
\end{center}
\caption{Frequency ratio, $\omega_{n=1}/\omega_{n=0}$, of the kink
mode vs $\alpha$ for $\eta=0.01$ and four different values of $\xi$
equal to 0.01 (red line), 0.1 (green line), 0.2 (purple line) and
0.3 (blue line)\textbf{,} in the case of $\alpha_{\rm e}=\alpha$.}
\end{figure}



\begin{figure}
\vspace*{1mm}
\begin{center}
\includegraphics[width=9.5cm]{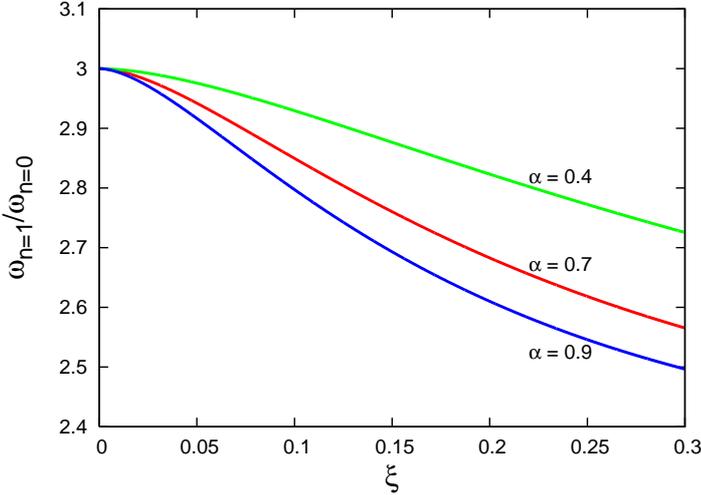}
\end{center}
\caption{Frequency ratio, $\omega_{n=1}/\omega_{n=0}$, of the kink mode vs $\xi$ for
$\eta=0.01$ and three different values of $\alpha$ equal to 0.4
(green line), 0.7 (red line) and 0.9 (blue line). Here $\alpha_{\rm e}=\alpha/\sqrt\eta$ is assumed.}
\end{figure}

\begin{figure}
\vspace*{1mm}
\begin{center}
\includegraphics[width=9.5cm]{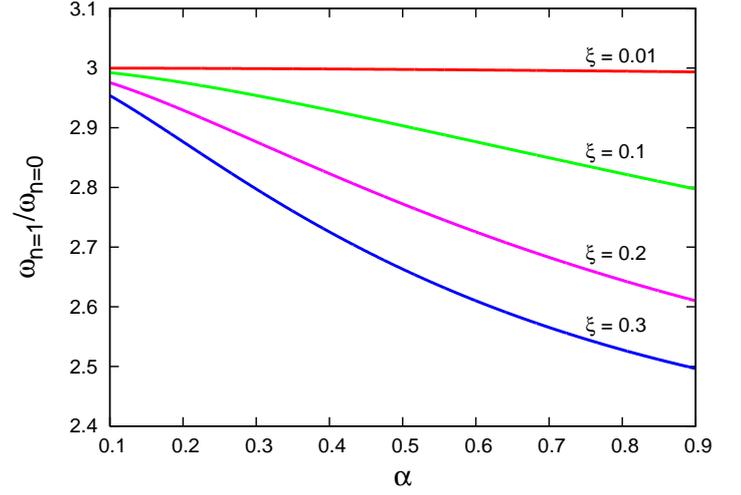}
\end{center}
\caption{Frequency ratio, $\omega_{n=1}/\omega_{n=0}$, of the kink mode vs $\alpha$ for
$\eta=0.01$ and four different values of $\xi$ equal to 0.01 (red
line), 0.1 (green line), 0.2 (purple line) and 0.3 (blue line). Here $\alpha_{\rm e}=\alpha/\sqrt\eta$ is assumed.}
\end{figure}

\subsection{Numerical solution} 
\label{S-equations}

We solved Eq.~(\ref{trans}) numerically for sausage ($m=0$) and kink
($m=1$) modes in the prominence conditions.
The inhomogeneity parameter of external density, $\alpha_{\rm e}$, may affect the wave frequencies. Therefore, we solve the dispersion equation~(\ref{trans}) for two different values of $\alpha_{\rm e}$. First, we suppose that $\alpha_{\rm e}=\alpha$, so the plasma density inside and outside has same behaviour. Then we consider $\alpha_{\rm e}=\alpha/\sqrt{\eta}>\alpha$ and show how the stronger external inhomogeneity of plasma density influences the obtained results.

Figure 2 shows the dependence of frequency of first (which often is called as fundamental) and second harmonics on the ratio
of the tube width to length, $\xi=a/L$, for $\alpha=0.7$. The plots are
similar to that of Edwin and Roberts (\cite{Edwin1983}). The fundamental kink
mode is expressed by horizontal lines in the upper plot. It is seen in Figure 2 that the frequencies of $n=0$ and $n=1$ harmonics of fundamental kink mode differs by a factor of 3. However, the two curves are almost identical for higher harmonics which correspond to higher zeros of Bessel functions. These modes correspond to the propagation of waves almost along radial direction, which means that their frequency is determined by the tube radius $a$ not by the tube length $L$ (similar result was found recently by Nakariakov et al. \cite{Nakariakov12} for sausage modes of a uniform cylinder). For example, the normalized frequency of second kink mode at $\xi=0.1$ is $\sim \!\! 45$, which shows that the corresponding wavelength is of order of $a$. It means that the frequencies of $n=0$ and $n=1$ modes are determined by the tube radius $a$, therefore they have similar values.

Figure~3 shows the dependence of a fundamental kink mode frequency on
the inhomogeneity parameter $\alpha$. The curves corresponding to the
different values of $\xi$ are practically identical for both, first and
second harmonics, therefore we provide only curves for one value.
On the contrary, the frequency of a fundamental kink mode depends
linearly on the inhomogeneity parameter $\alpha$. This is in agreement with analytical formula, Eq.~(\ref{fund}).
It is also seen, that the ratio of second and first harmonic frequencies is near $3$ in thin and weakly
inhomogeneous tubes in completed agreement with Eq.~(\ref{ratio}).


The dependence of the frequency ratio of second and first harmonics
of a fundamental kink mode on the ratio of the tube width to length is
shown in Fig.~4 for different values of $\alpha$. It is seen that
the frequency ratio only slightly decreases for wider tubes. The
decrease is more pronounced for the tubes with stronger longitudinal
inhomogeneity, but it is still insignificant in general.

Figure~5 shows the dependence of the frequency ratio on the
inhomogeneity parameter $\alpha$ for different values of $\xi$. The
frequency ratio is almost constant in both, thin and thick tubes (see red and blue lines).

Now we solve the dispersion equation, Eq.~(\ref{trans}), for the $\alpha_{\rm e}=\alpha/\sqrt{\eta}$ case.



The dependence of the frequency ratio of second and first harmonics
of fundamental kink mode on the ratio of tube width to length is shown in Fig.~6 for different
values of $\alpha$. It is seen that the frequency ratio
significantly decreases for wider tubes. The decrease is more
important for the tubes with strong longitudinal inhomogeneity.

Figure~7 shows the dependence of the frequency ratio on the inhomogeneity parameter
$\alpha$ for different values of $\xi$. The frequency
ratio is almost constant in thin tubes (see red line), but it significantly depends on $\alpha$ for wider tubes.

Thus, the numerical simulations of the dispersion relation, Eq.~(\ref{trans}), show that the frequency of a fundamental kink mode significantly depends on the inhomogeneity parameter, $\alpha$. On the contrary, the frequency ratio of second and first harmonics does not show significant dependence on density inhomogeneity when $\alpha_{\rm e}=\alpha$. However, when $\alpha_{\rm e}$ increases then the frequency ratio significantly depends on the this parameter for wider tubes. There is a good agreement between numerical solutions and analytical formulas, Eqs.~(\ref{fund}) and (\ref{ratio}).  Both, frequency and the frequency ratio do not depend significantly on the ratio of the tube width to length for $\alpha_{\rm e}=\alpha$. Therefore, the thin flux tube limit is a good approximation for fundamental kink waves as it is obtained by previous studies (Dymova and Ruderman \cite{Dymova2005}, D\'iaz et al. \cite{Diaz2010}) if the external density inhomogeneity is similar or smaller than the internal one.
Stronger density inhomogeneity in the external medium significantly alter the frequency ratio for wider tubes, therefore the thin flux tube approximation can be used with caution in this case. However, external plasma is usually hotter than the plasma inside prominence threads, which means that the external density inhomogeneity should be smaller than the internal one. Then it justifies the applicability of the thin flux tube approximation in prominence seismology.

\section{Discussion and conclusion} 
\label{S-Conclusion}

The period ratio of second and first harmonics is near 2 in weakly inhomogeneous coronal loops, however it is quite different in the case of prominences. Considering piecewise profiles of density concentration at the tube midpoint, several authors found that the ratio tends to 3. First, Joarder and Roberts (\cite{Joarder1992b}) found that the asymptotic frequency of internal kink mode is proportional to $2n+1$ (see Eq.~(21) in the paper), which yields that the period ratio of first and second harmonics of kink mode is 3 in the prominence case.
Second, one can see in Figs.~4 and 5 in Dymova and Ruderman (\cite{Dymova2005}) that the period ratio tends to 3 when $W/L$ approaches to $0.2$, where $W$ is the half length of thread and $L$ is the half length of the tube itself. Third, Fig.~6 in D\'iaz et al. (\cite{Diaz2010}) shows that the period ratio of first and second harmonics of kink mode tends to 3 for $W/L > 0.1$. Joarder and Roberts (\cite{Joarder1992b}), Dymova and Ruderman (\cite{Dymova2005}) and D\'iaz et al. (\cite{Diaz2010}) did not explicitly note this fact and consequently did not explain why the period ratio has tendency towards $3$ in prominences.

In this paper, we showed that the tendency of period ratio towards $3$ is the result of analogy between prominence oscillations and oscillations of quantum harmonic oscillator. The density enhancement at the midpoint of longer tube, which appears as a prominence thread, plays a role of potential energy. We used the method of separation of variables and solve the Sturm-Liouville problem of bounded oscillations in a prominence thread with longitudinally inhomogeneous density of parabolic profile. We found that the spatial variation of total pressure along the tube axis is governed by the stationary Schr\"{o}dinger equation, where the term with the longitudinal inhomogeneity of density stands for the potential energy. The equation is transformed into the equation of parabolic cylinder for the parabolic profile of the density. Consequently, the solutions are found in terms of Hermite polynomials, which are a set of orthogonal polynomials over the domain $(-\infty, \infty)$. Therefore, the
solutions form an orthogonal basis of the Hilbert space. The solutions are bounded in infinity, therefore the oscillations are trapped inside the thread and hence are localised much higher than the footpoints. Then, using the continuity of velocity and total pressure at the tube surface,
we derived the transcendental dispersion equation for MHD
oscillations in terms of Bessel functions. The dispersion equation
is solved analytically in thin flux tube approximation for kink
waves. It is obtained that the normalized frequency of a fundamental
kink mode depends on the inhomogeneity parameter $\alpha$ as
${\omega_n}\sim \alpha (2n+1)$, where $n$ is the longitudinal wave mode. This expression shows that the stronger inhomogeneity
leads to the higher frequency of kink waves. However, the ratio
of second and first harmonics of kink waves does not significantly depend on $\alpha$ in the thin tube approximation and it tends to 3
as it is suggested by the analogy with quantum mechanics.


We solved the dispersion equation numerically for kink and sausage
waves. The numerical solutions agree with the analytical
estimations. We found that the frequency ratio of second and first
harmonics of kink waves tends to 3 in thin flux tubes. The ratio depends slightly on the width to length
ratio of the tube (Fig.~4). The wider tubes lead to bit smaller
ratio of the frequencies when the inhomogeneity parameter inside and outside the tube is the same. Therefore, the thin flux tube
approximation is a good approximation for the estimation of the
ratio in this case. However, when the external inhomogeneity of plasma density is stronger then
the ratio significantly depends on the inhomogeneity for wider tubes and the thin tube approximation may fail.
Numerical solution of the dispersion equation shows that the frequency ratio
slightly depends on the longitudinal inhomogeneity and decreased for
stronger $\alpha$ (Fig.~5) for thin tubes, which is in good agreement with analytical formulas.

The calculation in this paper is performed for straight tubes, while
the curvature may significantly affect the oscillation spectrum
(Selwa et al. \cite{Selwa2005}, D\'iaz et al. \cite{Diaz2006}). It would be also interesting to study the
case of parabolic longitudinal inhomogeneity, when the denser plasma is
located near the tube ends. In this case, the tube is analog of
coronal loops rather than of prominence threads. The both problems
should be studied in the future.

\begin{acknowledgements}
The work was supported by the Austrian Fonds zur F\"orderung der Wissenschaftlichen Forschung (project P26181-N27) and by the European FP7-PEOPLE-2010-IRSES-269299 project- SOLSPANET. The work of TZ was also supported by Shota
Rustaveli National Science Foundation grant DI/14/6-310/12. The work of IZh was supported by the Bulgarian
Science Fund under project CSTC/INDIA 01/7.
\end{acknowledgements}


\end{document}